\documentclass[10pt,
							 twocolumn,
							 superscriptaddress,
							 english,
							 prl,
							 showpacs,
							 floatfix,
							 aps
							]{revtex4-1}
\usepackage[utf8]{inputenc}
\usepackage{amsmath}
\usepackage{amssymb}
\usepackage{graphicx}
\usepackage{xspace}

\makeatletter
\@ifundefined{textcolor}{}
{%
 \definecolor{BLACK}{gray}{0}
 \definecolor{WHITE}{gray}{1}
 \definecolor{RED}{rgb}{1,0,0}
 \definecolor{GREEN}{rgb}{0,1,0}
 \definecolor{BLUE}{rgb}{0,0,1}
 \definecolor{CYAN}{cmyk}{1,0,0,0}
 \definecolor{MAGENTA}{cmyk}{0,1,0,0}
 \definecolor{YELLOW}{cmyk}{0,0,1,0}
}

\usepackage{soul}
\usepackage{braket}
\usepackage[backref=none,
bookmarksnumbered=true,
bookmarks=true,
bookmarksopen=true,
colorlinks=true,
citecolor=blue,
linkcolor=blue,
anchorcolor=green,
urlcolor=blue,unicode=false]{hyperref}

\let\baraccent=\= 
\renewcommand{\=}[1]{\stackrel{#1}{=}} 

\newcommand{\unitspace}{~}

\newcommand{\didv}{\ensuremath{\mathrm{d}I/\mathrm{d}V}\xspace}

\newcommand{\Fig}[1]{Fig.\unitspace\ref{fig:#1}}
\newcommand{\Figure}[1]{Figure\unitspace\ref{fig:#1}}

\DeclareMathOperator{\umueV}{\unitspace\mathrm{\mu eV}}

\makeatother

\usepackage{babel}

\begin{document}

\title{Interfering tunneling paths through magnetic molecules on superconductors:\\ 
Asymmetries of Kondo and Yu-Shiba-Rusinov resonances}

\author{La\"etitia Farinacci}
\affiliation{\mbox{Fachbereich Physik, Freie Universit\"at Berlin, Arnimallee 14, 14195 Berlin, Germany}}

\author{Gelavizh Ahmadi}
\affiliation{\mbox{Fachbereich Physik, Freie Universit\"at Berlin, Arnimallee 14, 14195 Berlin, Germany}}

\author{Michael Ruby}
\affiliation{\mbox{Fachbereich Physik, Freie Universit\"at Berlin, Arnimallee 14, 14195 Berlin, Germany}}

\author{Ga\"el Reecht}
\affiliation{\mbox{Fachbereich Physik, Freie Universit\"at Berlin, Arnimallee 14, 14195 Berlin, Germany}}

\author{Benjamin W. Heinrich}
\affiliation{\mbox{Fachbereich Physik, Freie Universit\"at Berlin, Arnimallee 14, 14195 Berlin, Germany}}

\author{Constantin Czekelius}
\affiliation{Institut f\"ur Organische Chemie und Makromolekulare Chemie,
 Heinrich-Heine-Universit\"at D\"usseldorf, Universit\"atsstrasse 1, 40225 D\"usseldorf, Germany}

\author{Felix von Oppen}
\affiliation{\mbox{Dahlem Center for Complex Quantum Systems and Fachbereich Physik, Freie Universit\"at Berlin, 14195 Berlin, Germany}}

\author{Katharina J. Franke}
\affiliation{\mbox{Fachbereich Physik, Freie Universit\"at Berlin, Arnimallee 14, 14195 Berlin, Germany}}

\date{\today}

\begin{abstract}

Magnetic adsorbates on superconductors induce a Kondo resonance outside and Yu-Shiba-Rusinov (YSR) bound states inside the superconducting energy gap. When probed by scanning tunneling spectroscopy, the associated differential-conductance spectra frequently exhibit characteristic bias-voltage asymmetries. Here, we observe correlated variations of Kondo and YSR asymmetries across an Fe-porphyrin molecule adsorbed on Pb(111). We show that both asymmetries originate in interfering tunneling paths via a spin-carrying orbital and the highest occupied molecular orbital (HOMO). Strong evidence for this model comes from nodal planes of the HOMO, where tunneling reveals symmetric Kondo and YSR resonances. Our results establish an important mechanism for the asymmetries of Kondo and YSR lineshapes. 

\end{abstract}

\pacs{%
			} 
\maketitle 


Exchange coupling a magnetic adsorbate to a metallic substrate induces a many-body Kondo resonance at the Fermi level. 
Already initial observations of the Kondo effect of single magnetic adatoms by scanning tunneling spectroscopy revealed asymmetric Fano lineshapes \cite{Madhavan1998,Li1998}. These have been interpreted as arising from interference between tunneling via the adatom and direct tunneling into the continuum of electronic states of the metallic substrate \cite{Schiller2000,Ujsaghy2000,Ternes2008,Figgins2010}. On superconducting substrates, the exchange coupling also binds a pair of Yu-Shiba-Rusinov (YSR) states inside the superconducting energy gap \cite{Yu1965,Shiba1968,Rusinov1969}, which coexist with the Kondo resonance outside of the gap \cite{Matsuura1977,Sakai1993,Zitko2015}. Corresponding tunneling-spectroscopy resonances appear for both bias polarities with identical energies but asymmetric strengths \cite{Yazdani1997, Ji2008, Franke2011}. This asymmetry is commonly associated with differences between the electron and hole components of the YSR wavefunction due to breaking of particle-hole symmetry by additional potential scattering  \cite{Yu1965,Shiba1968,Rusinov1969,Flatte1997, Salkola1997,Balatsky2006}.

Within these conventional interpretations, the asymmetries of the Kondo and YSR resonances seem unrelated. Surprisingly, our measurements reveal striking correlations between these asymmetries when tracking them as a function of tip position above a Fe-5,10,15,20-tetra-pyridyl-porphyrin molecule (FeTPyP) with its spin moment derived from the Fe center and large extent over the organic ligand \cite{Liu2017, Verdu2017, Rolf2018}. Placing the molecules on a Pb(111) substrate, we investigate the YSR states in the presence of superconductivity and the Kondo resonance while quenching superconductivity by a magnetic field. This allows us to record YSR and Kondo spectra with the same tip across the very same molecule. 

These correlations suggest that the origins of the Kondo and YSR asymmetries have to be reconsidered for our system. Recent theoretical work by Frank and Jacob suggested that the Kondo asymmetry of magnetic adatoms originates from a parallel tunneling path through $s$ or $p$ orbitals within the atom itself \cite{Frank2015} (see also \cite{Fernandez2020}). While their predictions have not yet been validated by experiment, experiments investigated spatial variations of the Fano lineshape in (metal-organic) molecules \cite{Perera2010,Requist2014,Wang2015, Meyer2016}. YSR states were observed in several molecular systems, though without the aim of spatially resolving their asymmetries \cite{Franke2011, Hatter2015,Kezilebieke2018, Farinacci2018, Malavolti2018, Brand2018, Etzkorn2018, Kezilebieke2019}. We find that parallel tunneling through two molecular orbitals
naturally explains the correlations between the Kondo and YSR resonances, consistent with and extending the basic picture proposed in Ref.\ \cite{Frank2015}. 

\begin{figure*}[tb]
	\includegraphics[width=\textwidth]{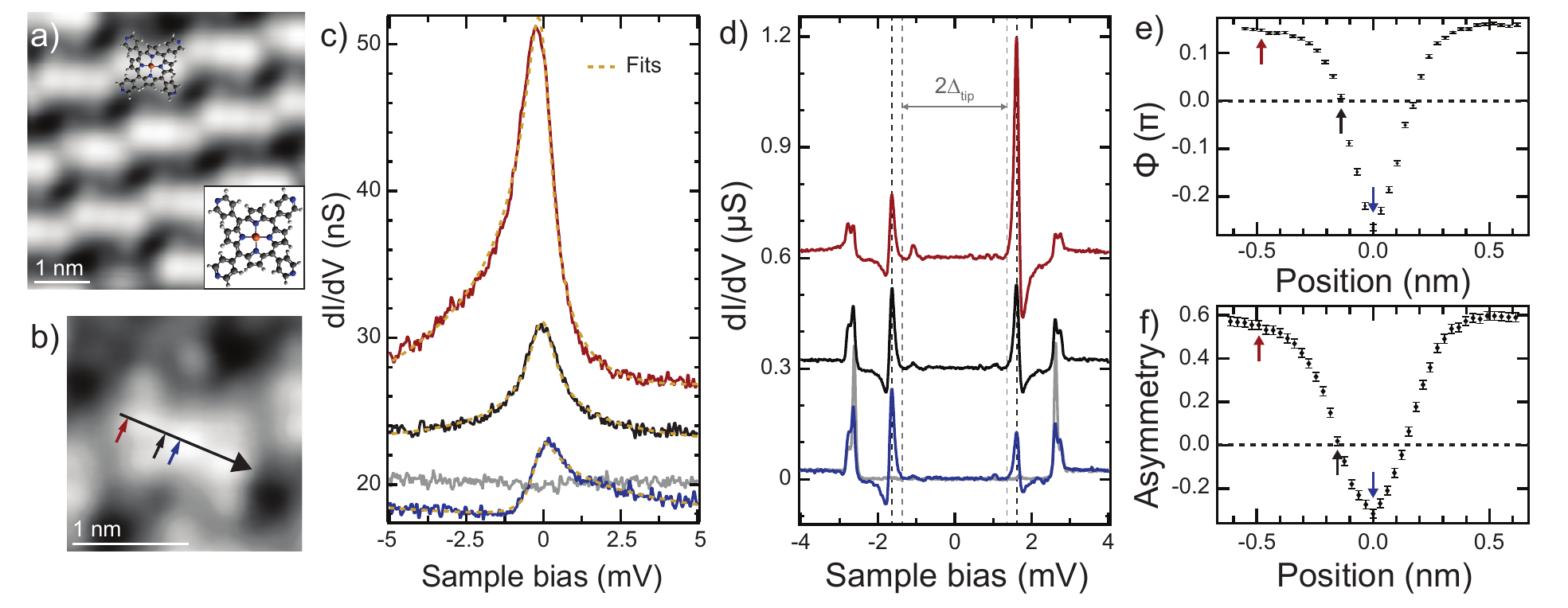}
	\caption{(a) Topography of molecular island ($V=-800$\,mV, $I=50$\,pA). Inset: molecular structure of the FeTPyP molecule. (b) Closeup of one molecule with the color-coded locations of the spectra in (c,d) marked by arrows ($V=5$\,mV, $I=100$\,pA). (c) \didv spectra of the Kondo resonance, measured in a magnetic field of $B=1.5$\,T. Gray spectrum taken on the bare Pb(111) substrate, colored spectra taken at locations marked in (b) (feedback opened at $V=5$\,mV, $I=100$\,pA, lock-in $V_\mathrm{rms}=100$\,$\mu$eV). Spectra are offset for clarity. Fits of the resonances to the Frota-Fano function (\ref{FFls}) are shown as dashed brown lines. (d) \didv spectra recorded with superconducting tip and substrate in the absence of the $B$-field. The YSR states (marked by dashed black lines) appear as a pair of resonances symmetric in energy, but asymmetric in intensity. The tip gap is indicated by gray dashed lines. Peaks below this energy are due to thermal excitations of the YSR states. Gray spectrum exhibiting BCS coherence peaks taken on the bare substrate, colored spectra taken at locations marked in (b). Spectra are offset for clarity (feedback opened at $V=5$\,mV, $I=100$\,pA, lock-in $V_\mathrm{rms}=\,15$\,$\mu$eV). (e) Asymmetry parameter $\phi$ deduced from a Frota-Fano fit of the Kondo spectra along the line indicated in (b) (zero marking the molecule's center). Error bars represent the fit error. (f) Asymmetry $(H_+-H_-)/(H_++H_-)$ of the YSR resonances along the line indicated in (b). Peak heights $H_\pm$ are determined by a search algorithm for local maxima with an estimated error of 0.02 due to noise.
	}
	\label{fig:1}
\end{figure*}

FeTPyP-Cl molecules were evaporated from a Knudsen cell heated to $400^{\circ}$C onto a Pb(111) surface held at $40^{\circ}$C. This leads to the formation of densely-packed islands, with scanning-tunneling-microscope (STM) images of the molecules displaying a bright elongated shape (\Fig{1}a); for details, see Supplementary Material (SM) \cite{SM}. This appearance is attributed to a saddle-shaped conformation with two upward- and two downward-tilting pyrrole groups, reflecting competition between molecule--substrate interactions and intramolecular steric hindrance \cite{Auwarter2007, Chen2017, Liu2017, Verdu2017, Rolf2018}. The absence of a central protrusion shows that the central Cl ligand detaches from all molecules during adsorption \cite{Rolf2018}. This de-chlorination reduces the Fe oxidation state from +3 to +2, and the molecule is expected to lie on the surface in an integer spin state \cite{Heinrich2015,Liu2017,Verdu2017,Rolf2018}. STM images reveal two distinct types of molecules. Here, we focus on the majority type (\Fig{1}a), which exhibits both Kondo and YSR resonances. Spectra taken on the other type (not shown) exhibit inelastic spin excitations reflecting different interaction strengths with the substrate.

To probe the Kondo resonance and its spatial variation along the FeTPyP molecule, we record \didv spectra at $T=1.1$\,K in a magnetic field of 1.5\,T which quenches superconductivity in both substrate and tip. We find a strong resonance with a shoulder at negative bias voltages on the upper pyrrole (\Fig{1}b,c red line). As the tip position moves towards the center of the molecule, the peak height reduces and eventually inverts its bias asymmetry at the Fe center (\Fig{1}b,c blue line). These resonances are well fitted by Fano-Frota lineshapes \cite{Frota1986,Frota1992} 
\begin{equation}
\frac{\mathrm{d}I}{\mathrm{d}V} = A \textrm{Re}\left({e^{i\phi}\sqrt{\frac{i\Gamma/2}{eV-\omega_0+i\Gamma/2}}}\right)+B,
\label{FFls}
\end{equation}
where $A$ is the amplitude of the resonance at energy $\omega_0$, $\Gamma$ a width parameter,  $\phi$ a phase characterizing the asymmetry, and $B$ a bias-independent background. We find the same width of $(0.77\pm 0.03)$\,meV for all spectra along the molecule, while the asymmetry $\phi$ varies continuously and eventually inverts (\Fig{1}e). (Fits are shown as dashed lines in \Fig{1}c; for a complete set of spectra along the line indicated in \Fig{1}b, see SM). The narrow lineshapes and characteristic asymmetries suggest that these are Kondo resonances probed by interfering tunneling paths \cite{Rosch2003,Pruser2011, Frank2015, Gruber2018}. An observed splitting of the resonance at larger magnetic fields further confirms the magnetic origin (see SM). 

In addition to the Kondo resonance for a normal-state substrate, the exchange coupling also induces YSR bound states for a superconducting substrate. We investigate their dependence on tip position in the absence of the external magnetic field. Then, the Pb tip is also superconducting, increasing our energy resolution beyond the Fermi-Dirac temperature limit and shifting all features in the \didv spectra by the tip's energy gap $\Delta = 1.35$~mV \cite{RubyPb2015}. A spectrum taken on the bare Pb(111) substrate is shown in gray in \Fig{1}c. The superconducting energy gap is flanked by two pairs of coherence peaks at $\pm 2.65$\,mV and $\pm 2.8$\,mV, reflecting the two-band character of superconductivity in Pb \cite{RubyPb2015}. 

\didv spectra on the FeTPyP molecule show a pair of YSR resonances at $\pm 1.6$~mV inside the superconducting energy gap. Consistent with the observation of a constant Kondo width, their energies are independent of tip position above the molecule, indicating that the exchange coupling between substrate and molecule induces a single pair of YSR states. Unlike the energy, the heights $H_\pm$ of the YSR resonances are different at positive and negative bias voltages. In \Fig{1}f, we plot the measured asymmetry $(H_+-H_-)/(H_++H_-)$ vs.\ tip position. Comparison with the asymmetry variations of the Kondo resonance in \Fig{1}g reveals striking similarities. In particular, we observe a particle-hole symmetric YSR spectrum exactly where the Kondo peak exhibits a symmetric Frota peak (black spectra in \Fig{1}c,d). Comparing spectra on both sides of this symmetry point, we observe an inversion of the asymmetry of the YSR states, which parallels the asymmetry inversion of the Kondo resonance. 

\begin{figure}[tb]
	\includegraphics[width=\columnwidth]{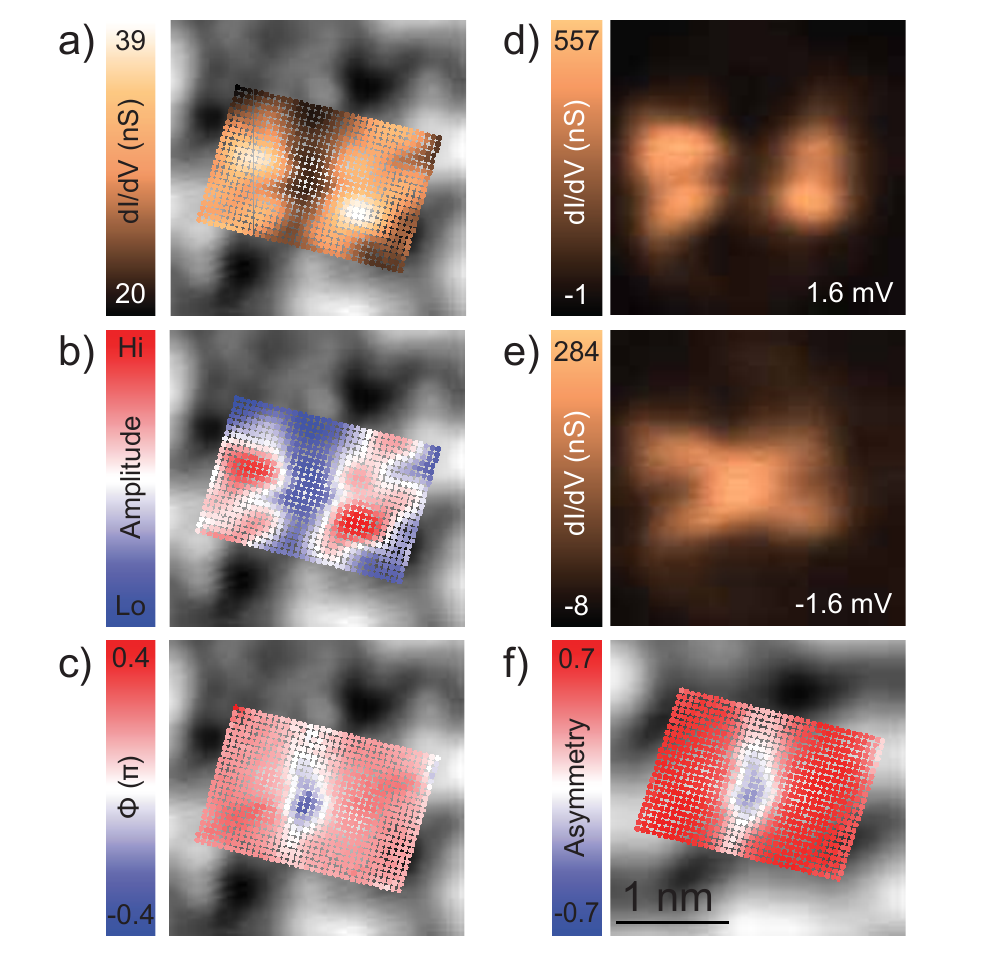}
	\caption{ (a-c) Topography of the molecule (background black-white images, $V=5$\,mV, $I=100$\,pA) with superimposed color plots extracted from a grid of spectra (feedback opened at $V=5$\,mV, $I=100$\,pA, signal modulated with $V_\mathrm{rms}=100\,\mu$eV). Color plots show (a) the \didv signal at $V=0$\,mV, (b) the amplitude $A$, and (c) the asymmetry parameter $\phi$ extracted from the Fano-Frota fits of the Kondo resonance to Eq.\ (\ref{FFls}). (d) \didv maps of the YSR states recorded along the topography contour given by $V=5$\,mV, $I=100$\,pA at positive bias voltage $V=1.6$\,mV and (e) at negative bias voltage $V=-1.6$\,mV ($V_\mathrm{rms}=15$\,$\mu$eV). (f) Asymmetry of the YSR state extracted from a grid of spectra taken (feedback opened at $V=5$\,mV, $I=100$\,pA, signal modulated with $V_\mathrm{rms}=15\,\mu$eV).Background black-white topography image recorded with $V=-800$\,mV, $I=50$\,pA. All images are recorded above the same area.}
	\label{fig:2}
\end{figure}

To further corroborate these correlations, we map out the Kondo and YSR resonances across the entire molecule. \Figure{2}a presents an STM image (grayscale) of a single FeTPyP molecule superimposed with zero-bias \didv intensities extracted from a densely spaced grid of \didv spectra. The four-lobe structure shows that the Kondo intensity is largest on the molecular ligand. We extract the local amplitude $A$ (\Fig{2}b) and phase parameter $\phi$ (\Fig{2}c) of the Kondo resonance across the entire molecule by fitting all \didv spectra to the Fano-Frota lineshape (\ref{FFls}). The amplitude naturally follows the \didv intensity at zero bias. Consistent with the line profile in \Fig{1}e, the asymmetry exhibits reversed signs above Fe atom and ligand. 

The YSR intensities at positive and negative bias voltages are shown in \Fig{2}d and e. At positive bias voltage, the YSR intensity has its maxima on the ligand, resembling the map of the amplitude of the Kondo resonance. At negative bias voltage, the YSR intensity is largest  at the Fe center. These intensity distributions are reflected in the map of the asymmetry shown in \Fig{2}f. In agreement with the line profile in \Fig{1}f, we find reversed asymmetries at the Fe center and the ligand. This map of the YSR asymmetry is remarkably similar to the map of the Kondo asymmetry in \Fig{2}c. These experimental observations strongly suggest a common origin of the Kondo and YSR asymmetries. 

\begin{figure}[tb]
	\includegraphics[width=\columnwidth]{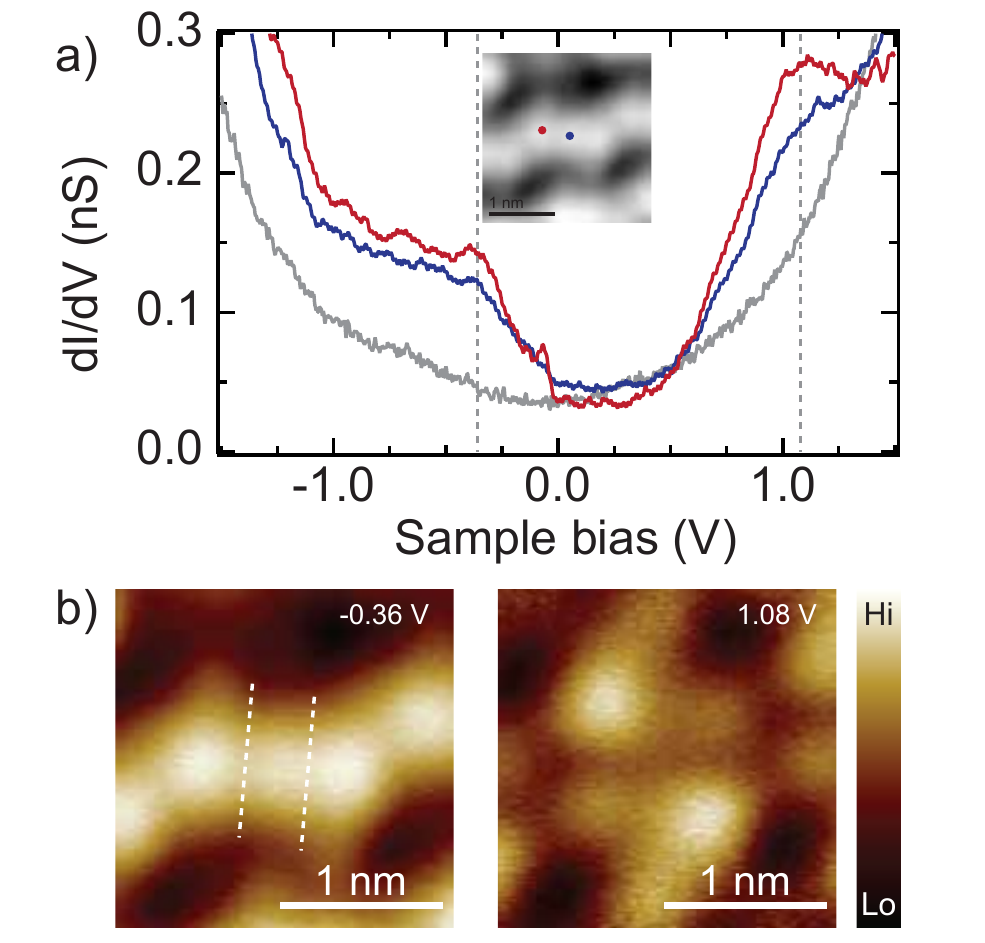}
	\caption{(a) \didv spectra taken on FeTPyP molecule (color-coded locations marked in inset) and on the bare Pb substrate (gray) (feedback opened at $V=2$\,V, $I=500$\,pA, and signal modulated with V$_\mathrm{rms}=10$\,mV). Dashed lines indicate locations of resonances identified as HOMO ($V\simeq -360$\,mV) and LUMO ($V\simeq -1.08$\,V). (b) Iso-density-of-states maps of molecule recorded at bias voltages as indicated in the top-right corners and corresponding to HOMO (left) and LUMO (right) resonances (signal modulated with V$_\mathrm{rms}=2$\,mV and regulated at a conductance of $2.06$\,nS). Nodal planes are indicated by dashed lines.}
	\label{fig:3}
\end{figure}

The observations of a constant Kondo-resonance width and YSR energy indicate that tunneling probes the same excitations across the entire molecule. While details of the spin state are not central to our model, the magnetic-field splitting of the Kondo resonance suggests an $S=1$ impurity spin, carried by hybrid orbitals of the ligand-field-split $d$ states and the organic ligand \cite{Verdu2017} (see also SM). The exchange coupling of one of the spin-carrying orbitals leads to the Kondo resonance \cite{Minamitami2015} and the pair of YSR states, both of which can be probed with varying intensity across the entire molecule.

While virtual excitations of this orbital generate both exchange and potential scattering of substrate electrons (as described by a Schrieffer-Wolff transformation), the pronounced Kondo correlations imply that exchange coupling is strongly dominant at low energies. Without additional tunneling channels, one thus expects not only a symmetric Kondo lineshape, but also symmetric YSR resonances, reflecting approximately equal electron and hole YSR wavefunctions. 

\begin{figure}[t]
	\includegraphics[width=\columnwidth]{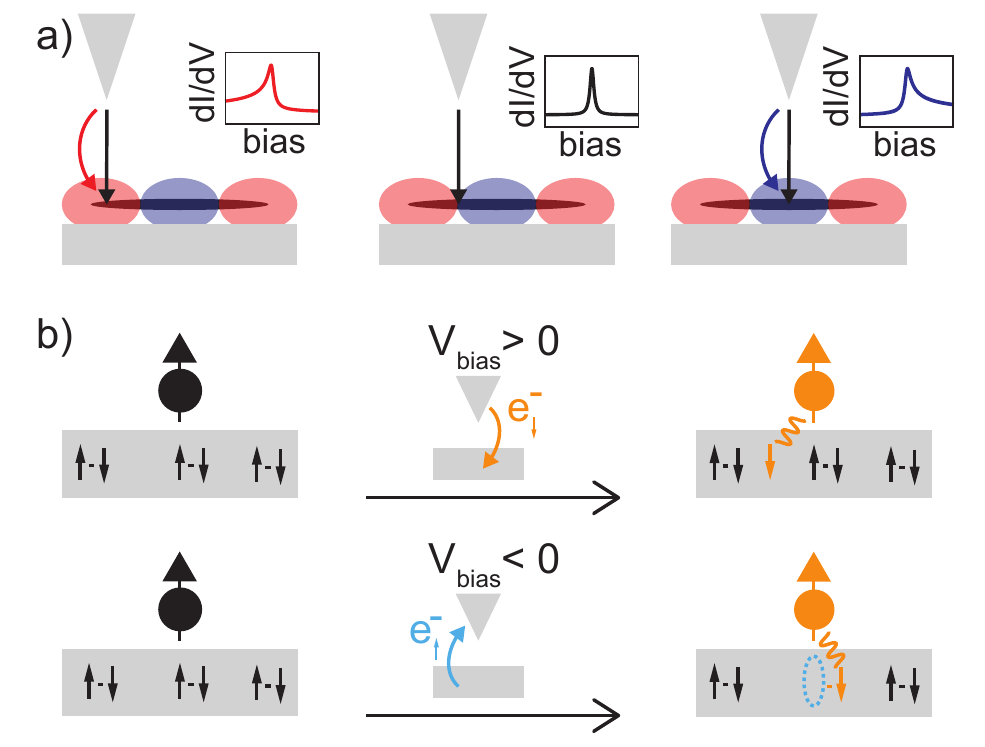}
	\caption{(a) Schematic HOMO wavefunction (red/blue) and spin-carrying molecular orbital (gray) on a surface for different tip positions, with the corresponding Kondo lineshapes in the insets. At the HOMO nodal plane, the lineshape is symmetric. (b) Tunneling into the superconducting substrate at positive (top) and negative bias (bottom) populating the YSR excitation (right) from an unscreened-spin ground state (left). Due to spin polarization of the YSR state and singlet Cooper pairing, the spin of electrons tunneling (into) from the substrate is (anti)parallel to the impurity spin.}
	\label{fig:4}
\end{figure} 

The asymmetries of the Kondo and YSR resonances point to a second tunneling path which exists in parallel to tunneling via this strongly exchange-coupled orbital. \didv spectra taken over a larger voltage range (\Fig{3}a) reveal a resonance centered at -360\,mV which we identify with the highest occupied molecular orbital (HOMO) and which extends towards the Fermi level. We associate the second tunneling path with this orbital. Its constant density-of-states map shows large intensity at the Fe center, separated by two nodal planes from maxima at the upper pyrroles of the molecular saddle (\Fig{3}b). These nodal planes of the HOMO wavefunction imply sign changes in the associated tunneling amplitude.

Strikingly, the nodal planes of the HOMO match with the locations where the Kondo and YSR spectra are symmetric and their asymmetries invert. This provides strong evidence for an interference model involving two molecular orbitals rather than parallel tunneling directly into the substrate (cp.\ Ref.\ \cite{Frank2015} for magnetic adatoms). For tip positions at the nodal planes, tunneling via the HOMO is strongly suppressed. Then, tunneling proceeds predominantly via the strongly exchange-coupled molecular orbital, and leads to symmetric Kondo and YSR lineshapes. Due to the sign change in the amplitude for tunneling via the HOMO, the two tunneling paths interfere with opposite relative sign on the two sides of the nodal plane, explaining the observed inversion of the asymmetry in the Kondo lineshape [see \Fig{4}a for a sketch and SM for a Green-function calculation relating this sign change to the asymmetric Kondo lineshape in Eq.\ (\ref{FFls})]. 

This model also provides a natural explanation for the YSR asymmetry and its correlations with the Kondo asymmetry. 
A shift of the YSR energy upon lifting the molecule further from the substrate using tip-molecule interactions implies that on the superconductor, the impurity spin is in the unscreened regime despite the sizable Kondo correlations \cite{Farinacci2018,SM}. Then, the ground state $|\mathrm{e}\rangle$ of the superconducting substrate has even fermion parity and tunneling excites the substrate into the odd-parity state $|\mathrm{o}\rangle = \gamma^\dagger_\epsilon |\mathrm{e}\rangle$, in which the Bogoliubov quasiparticle (with annihilation operator $\gamma_\epsilon$) associated with the YSR state is occupied. Due to the spin polarization of the YSR state \cite{Yu1965,Shiba1968,Rusinov1969,Flatte1997,Salkola1997,Kaladzhyan2016,Korber2018,Cornils2017}, the odd-parity state $|\mathrm{o}\rangle$ is only excited by electrons whose spin is antiparallel to the molecular spin, when tunneling from tip to substrate. Conversely, when tunneling from the substrate into the tip, the tunneling electron has parallel spin and leaves behind an excitation with antiparallel spin due to the spin-singlet nature of Cooper pairs (see \Fig{4}b for a sketch). This implies that the amplitude $\mp J_d$ for tunneling via the strongly exchange-coupled orbital has opposite signs for the two bias polarities, while the amplitude $V_{h}$ for tunneling via the HOMO is spin independent. A Fermi's golden rule calculation gives (see SM for details)
\begin{equation}
   \left.\frac{\mathrm{d}I}{\mathrm{d}V}\right|_\pm \propto |\langle {\rm o}|(\mp J_d S+V_{h})\mathcal{O}_\pm|{\rm e}\rangle|^2.
\end{equation}
In accordance with our physical arguments, $\cal{O}_+=\psi_\downarrow^\dagger$ at positive bias and  $\cal{O}_-=\psi_\uparrow$ at negative bias, where $\psi_{\sigma}$ annihilates a spin-$\sigma$ electron in the substrate. 
For equal weights of the electron and hole YSR wavefunctions, we have $|\langle {\rm o}|\psi_\downarrow^\dagger|{\rm e}\rangle|=|\langle {\rm o}|\psi_\uparrow|{\rm e}\rangle|$, and the YSR asymmetry reflects that the two tunneling paths interfere with opposite relative sign for the two bias polarities. The relative sign changes at the nodal plane due to $V_h$, explaining the asymmetry inversion of the YSR peaks and its correlation with the Kondo asymmetry. We note that the conclusions would remain qualitatively unchanged for a  screened impurity spin.

Additional orbitals could also contribute to the parallel channel, but we find the lowest unoccupied molecular orbital (LUMO) at a significantly higher energy of 1.08\,V. This is consistent with spatial maps of the LUMO (\Fig{3}c), which have their highest intensity at the four pyridine legs and unlike the HOMO do not match the symmetry properties of the Kondo or YSR asymmetries. 

In conclusion, we have shown that the asymmetry of the Kondo resonance across a magnetic molecule arises from interference between tunneling through different molecular orbitals, and need not involve direct tunneling into the substrate. Interference of the same tunneling paths also underlie a parallel asymmetry of the YSR resonances. Our observations also suggest an alternative understanding of YSR asymmetries, which have so far been interpreted in terms of different electron and hole wavefunctions due to potential scattering of substrate electrons on the magnetic impurity. While this interpretation suffices for magnetic adsorbates in the absence of additional tunneling paths \cite{Ruby2016, Choi2017}, we expect that our model will be widely relevant for interpreting tunneling spectroscopy experiments probing magnetic molecules on normal-metal and superconducting substrates. 

\begin{acknowledgments}
We thank J.\ I.\ Pascual, C.\ Rubio-Verd{\'{u}}, and Rok Zitko for discussions and sharing their manuscript on a similar porphyrin molecule prior to submission \cite{Verdu2020}. We gratefully acknowledge funding by the European Research Council under the Consolidator Grant ``NanoSpin'', and by Deutsche Forschungsgemeinschaft through project C03 of CRC 183. 
\end{acknowledgments}
  
\bibliographystyle{apsrev4-1}

%

\clearpage

\setcounter{figure}{0}
\setcounter{section}{0}
\setcounter{equation}{0}
\setcounter{table}{0}
\renewcommand{\theequation}{S\arabic{equation}}
\renewcommand{\thefigure}{S\arabic{figure}}
	\renewcommand{\thetable}{S\arabic{table}}%
	\setcounter{section}{1}
	\renewcommand{\thesection}{S\arabic{section}}%

\onecolumngrid

\renewcommand{\Fig}[1]{\mbox{Fig.\unitspace\ref{Sfig:#1}}}
\renewcommand{\Figure}[1]{\mbox{Figure\unitspace\ref{Sfig:#1}}}

\newcommand{\vsigma}{\mbox{\boldmath $\sigma$}}

\section*{\Large{Supplementary Material}}



\maketitle 

\section{Experimental details and additional data}

\subsection{Experimental details}
The Pb(111) substrate was cleaned by sputtering with Ne$^+$ ions at 0.9\,kV under ultra-high vacuum conditions. Annealing the sample at 430\,K leads to atomically flat and clean terraces. FeTPyP-Cl molecules were deposited from a Knudsen cell at 673\,K while the Pb sample was kept at 313\,K. The as-prepared sample was cooled down and transferred into the Joule-Thomson STM (by Specs). All experiments were carried out at a temperature of 1.1\,K. 

We used Pb-covered tips for all data shown in the main manuscript and SM. These were prepared by indenting the tip into the clean Pb substrate while applying a voltage of $V = 100$\,V. Their spectroscopic properties were tested on the bare Pb(111) substrate. The indentation procedure was repeated until the tip developed a bulk-like superconducting Bardeen-Cooper-Schrieffer (BCS) energy gap of $\Delta_\mathrm{tip} = 1.35$\,meV. Such a tip is of amorphous nature and exhibits a single gap, whereas the single-crystal substrate reveals the two-band superconducting properties pf Pb and exhibits two distinct BCS gaps with $\Delta_1 = 1.30$\,meV and $\Delta_2 = 1.44$\,meV  \cite{SRubyPb2015}. The sharp coherence peaks in the tip allow for an effective energy resolution beyond the thermal Fermi-Dirac limit. In essence, the superconducting gap of the tip leads to a shift of all sample features by $\Delta_\mathrm{tip}$. As the tip's quasi-particle density of states is symmetric in intensity around the Fermi level, the tip does not modify any asymmetries in the spectral properties of the sample. 

Differential conductance spectra were recorded using an external lock-in amplifier with a modulation frequency of $f=911$\,Hz. The modulation amplitudes are given in the corresponding figure captions.

We use different feedback control methods for recording \didv\ maps. In the ``constant-contour" mode, often also referred to as ``multi-paths", we first record a constant-current image with certain feedback parameters. These determine the height profile of the tip across the surface. This profile is subsequently used for scanning along the same path with the desired bias voltage while recording the \didv\ signal. The data is exactly the same (but much faster in its acquisition) as the \didv\ values extracted from a densely spaced grid of spectra recorded when opening the feedback at the set contour value. Figs.\ 2d,e have been acquired with this method. 

A bias voltage of 5\,mV was chosen when opening the feedback loop for the acquisition of all spectra in Fig.\ 1. A small bias voltage is ideal as it ensures that only a small range of electronic states contributes to the tunneling current. However, in the case of FeTPyP on Pb(111), the highest occupied molecular orbital (HOMO) extends across the Fermi level and we observe a non-negligible contribution to the tunneling current. Hence, the image is not only of topographic origin, but convolved with these states. For this reason, the STM image (Fig.\ 1b) does not only show the tilted pyridine legs, which are expected to be highest in topography \cite{SChen2017}, but also the three lobes of the HOMO resonance along the pyrrole saddle.

To map the iso-density of a particular orbital, the method of choice is to use the \didv\ signal itself for feedback control \cite{SReecht2017}. This procedure has been employed for the acquisition of the maps in Fig.\ 3b.

\subsection{Evolution of Kondo and YSR asymmetry along an FeTPyP molecule}

\begin{figure*}[tb]
	\includegraphics[width=\textwidth]{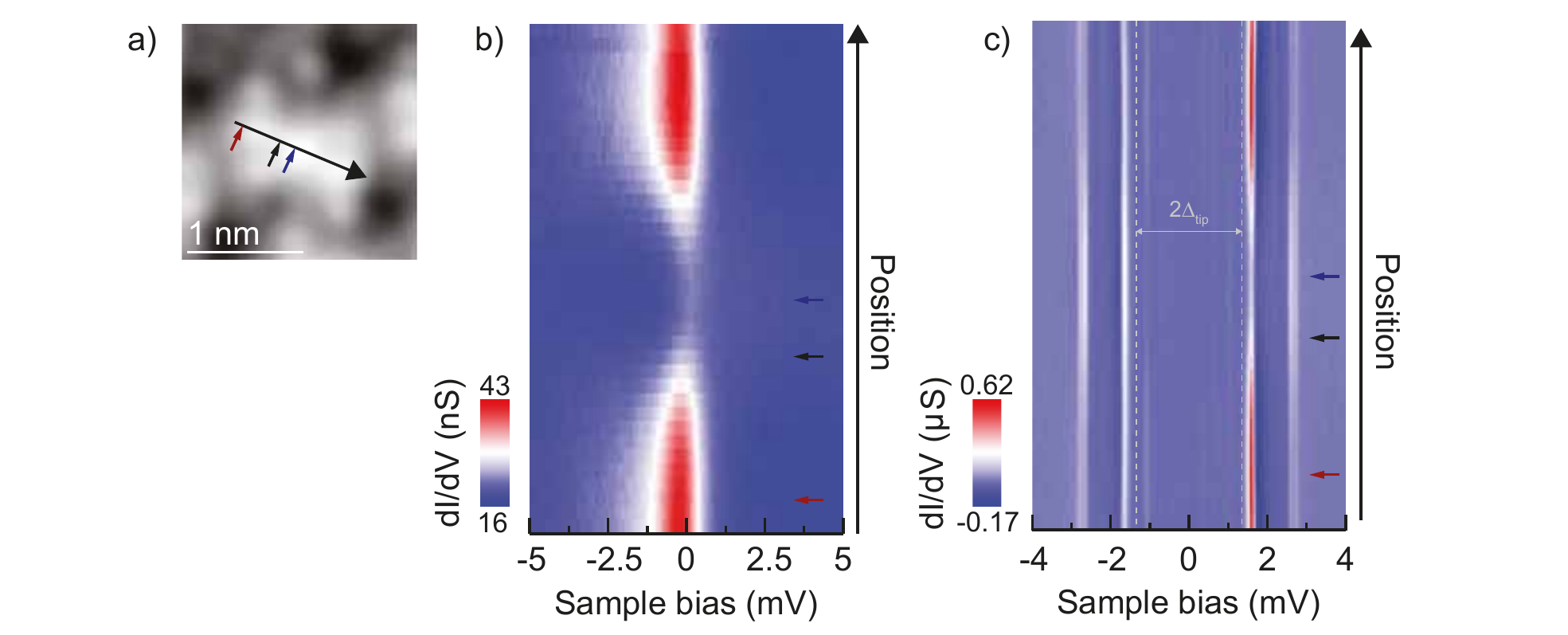}
	\caption{(a) Topography image of one molecule (same as in Fig.\ 1b) with an arrow indicating the line along which a set of Kondo and YSR spectra have been recorded  ($V=5$\,mV, $I=100$\,pA). The colored arrows mark the positions of the spectra in Fig.\ 1 and are also used as markers in (b,c). (b) Color plot of \didv spectra of the Kondo resonance measured in an external magnetic field of $B=1.5$\,T to quench superconductivity in substrate and tip (feedback opened at $V=5$\,mV, $I=100$\,pA and signal modulated with $V_\mathrm{rms}=100$\,$\mu$eV). These spectra have been used to extract the asymmetry shown in Fig.\ 1e. 
	(c) \didv spectra recorded with a superconducting tip and substrate in the absence of the $B$-field (feedback opened at $V=5$\,mV, $I=100$\,pA and signal modulated with $V_\mathrm{rms}=\,15$\,$\mu$eV). The superconducting energy gap of the tip is indicated by dashed lines. The YSR states appear as a pair of resonances symmetric in energy around the Fermi level, but asymmetric in intensity. These spectra have been used to extract the asymmetry shown in Fig.\ 1f. }
	\label{fig:S1}
\end{figure*}

In the main text, we show three representative spectra of the Kondo resonance and the YSR states measured along the pyrrole saddle of FeTPyP. In Fig.\,\ref{fig:S1}b,c, we show color plots of a series of densely spaced spectra, taken along the same line, for the Kondo and YSR states, respectively. These spectra were used to extract the asymmetry shown in Fig.\ 1e,f in the manuscript.

\subsection{Magnetic field dependence of Kondo resonance}

\begin{figure*}[htb]
	\includegraphics[width=\textwidth]{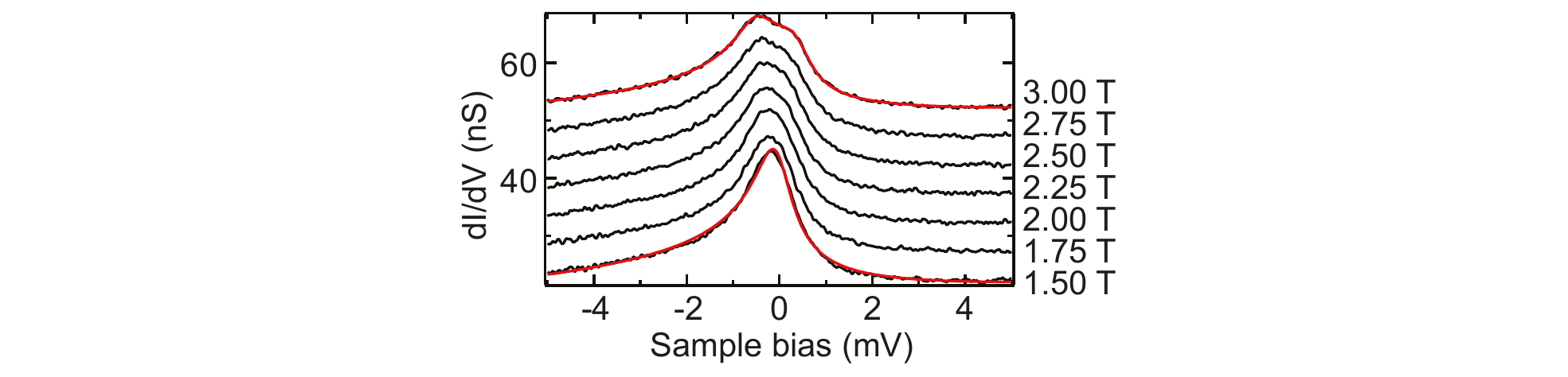}
	\caption{\didv spectra showing the evolution of the Kondo resonance with increasing magnetic field strength (feedback opened at $V=5$\,mV, $I=100$\,pA, modulation voltage $V_\mathrm{rms}=100\,\umueV$). The spectra are offset for clarity. Red lines are fits with one (at 1.5\,T) and two (at 3\,T) Fano-Frota functions. }
	\label{fig:S2}
\end{figure*}

To quench the superconducting state of substrate and tip, we applied an external magnetic field of $B=1.5$\,T perpendicular to the surface. The evolution of the Kondo resonance with increasing field strength is shown in Fig.\,\ref{fig:S2}. At 3\,T we observe a clear splitting of the Kondo resonance, which can be reproduced by two Fano-Frota functions shifted by $\pm 450 \mu$eV from the Fermi level. The observed splitting is larger than the $\pm 350 \mu$eV energy splitting expected for a spin $S=1/2$. We attribute this behavior to an underscreened $S=1$ Kondo system \cite{SColeman2003}. However, we note that our temperature and modulation broadening of $\sqrt{(2 V_\mathrm{rms}^2)+(3.5 k_B T)^2}=390 \umueV$ is of the same order of magnitude, preventing a clear assignment of the splitting.

We also note that at a field strength of 1.5\,T, we do not observe a splitting of the Kondo resonance, because the thermal broadening is still  much larger than the expected splitting. This justifies the fit with a single Fano-Frota function.

\subsection{Identification of YSR coupling regime}

\begin{figure*}[htb]
	\includegraphics[width=\textwidth]{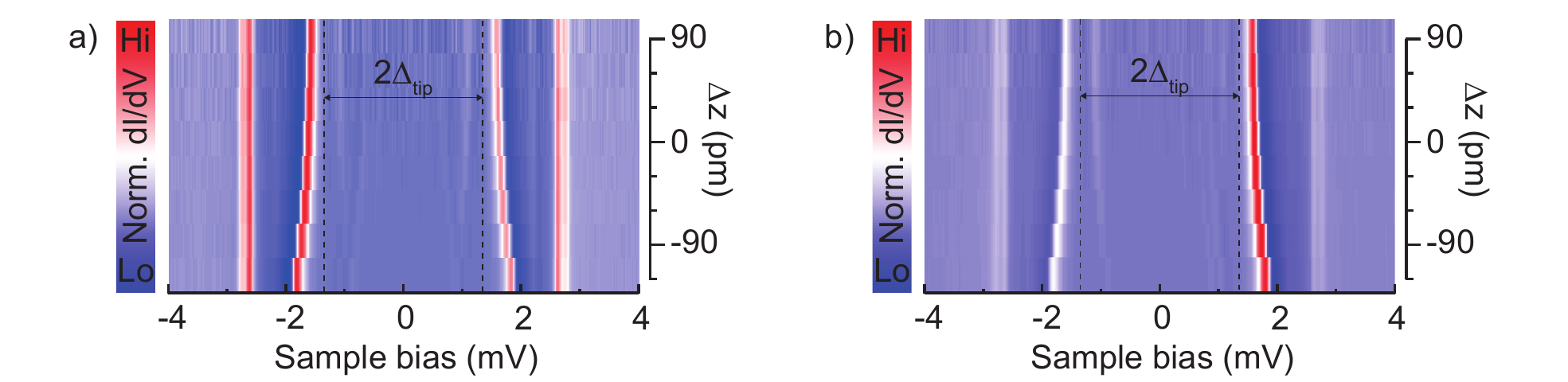}
	\caption{Normalized \didv spectra upon tip approach on (a) the Fe center and (b) upper pyrrole (feedback opened at $V=5$\,mV, $I=100$\,pA, and subsequent variation of tip height $z$, modulation voltage $V_\mathrm{rms}=15\,\umueV$. The superconducting energy gap of the tip is indicated by dashed lines. The YSR states shift toward the superconducting gap edge upon tip approach, indicating that the FeTPyP molecule lies in the unscreened spin regime. }
	\label{fig:S3}
\end{figure*}

The observation of a pair of YSR states reveals the exchange coupling to the substrate. However, this does not allow for the identification of the many-body ground state, i.e., for the assignment of the screened or unscreened spin state of the FeTPyP molecule (singlet and doublet many-body ground states, respectively). By approaching the STM tip toward the molecule, the attractive potential of the tip lifts the molecule from the substrate. Hence, we can investigate the shift of the YSR state with the associated decrease in the exchange coupling strength \cite{SFarinacci2018}. Figure\,\ref{fig:S3} reveals that the YSR states move towards the superconducting gap edge upon tip approach. This behavior shows that the molecule is in the unscreened regime. The narrow Kondo linewidth is in agreement with this assignment \cite{SFranke2011}.

\section{Theoretical considerations}

\subsection{Model}

We start by considering the spin state of Fe in FeTPyP. With the central Cl ligand detached, the molecule presumably is in oxidation state +2 and its square-planar ligand field splits the Fe $d$-shell into four sets. For an unoccupied topmost $d_{x^2-y^2}$ orbital, this results in a $S=1$ spin state. In order of ascending energy, one expects a doubly-occupied $d_{xy}$ orbital, degenerate $d_{xz}$ and $d_{yz}$ orbitals occupied by three electrons, and a singly-occupied $d_{z^2}$ orbital. Due to the largest overlap with the substrate, we expect the $d_{z^2}$ orbital to give rise to the Kondo effect, with the two unpaired electrons being Hund coupled \cite{SMinamitami2015}. This picture is consistent with the observed spin splitting of the Kondo resonance in a magnetic field (see Fig.\ \ref{fig:S2}) and with the observation of only a single Kondo temperature $T_K$, irrespective of tip location. We also identify only a single Yu-Shiba-Rusinov (YSR) state. In particular, there is no evidence for a second YSR state which splits off from the YSR peak or superconducting gap edge upon tip approach. These observations are consistent with an underscreened Kondo effect, in which the $S=1$ impurity spin is screened by a single channel of substrate electrons.  

The scanning-tunneling-microscopy (STM) data suggest that a second orbital is involved in determining the low-bias tunneling spectra, both with and without superconductivity. In particular, the resonance centered at $V = -360$\,mV weakly overlaps with the Fermi energy (see Fig.\ 3a of the main text) and thus constitutes a second tunneling channel. We identify this orbital with the HOMO. Mapping out this additional molecular orbital reveals nodal planes (see Fig.\ 3b of the main text) where its wavefunction changes sign. A central ingredient of our model is that tip-substrate tunneling via this second orbital includes a substantial potential-scattering amplitude. To bring out the essential physics, we will discuss our model in the following for the case that the HOMO is nonmagnetic in nature and contributes potential scattering only between tip and substrate. It should be understood, however, that this is not a necessary assumption for our model to explain the experimental results. If the HOMO is also magnetic, we merely have to assume that it does not develop strong Kondo correlations (as indeed suggested by the data which exhibit only a single Kondo temperature). In the absence of strong Kondo correlations, the exchange coupling has not yet flown to strong coupling at the relevant energy scales (as set by temperature and the superconducting gap) and tunneling between tip and substrate via the HOMO can include a substantial potential scattering amplitude in addition to the exchange-scattering contribution. In this situation, one finds qualitatively the same results as in the reduced model discussed below. In particular, we note that the exchange scattering amplitude contributed by tunneling via a magnetic HOMO would also change sign at the nodal plane so that the coincidence of symmetric Kondo and YSR peaks with the nodal plane is still be expected from such a modified model.

As a minimal model, we thus consider two molecular orbitals: (i) a singly-occupied orbital which derives from the Fe $d$ orbitals (but has nonzero amplitude both on the Fe center and the ligand as a result of hybridization) and develops a magnetic moment due to an onsite interaction $U$, and (ii) a doubly-occupied orbital which represents the HOMO and has negligible onsite interaction. The annihilation operator for an electron of spin $\sigma$ in the first, spin-carrying orbital of energy $\epsilon_d$ is denoted by $d_\sigma$. We perform a particle-hole transformation for the second (HOMO) orbital, resulting in a hole orbital of energy $\epsilon_h$ and hole-annihilation operator $h_\sigma$. (In the following, the hole nature of this orbital will be left implicit.) Then, the Hamiltonian becomes
\begin{eqnarray}
 H &=& \sum_{{\bf k}\sigma} \epsilon_{L\bf k} \psi^\dagger_{L{\bf k}\sigma} \psi^{\phantom\dagger}_{L{\bf k}\sigma} + \sum_{{\bf k}\sigma} \epsilon_{R\bf k} 
   \psi^\dagger_{R{\bf k}\sigma} \psi^{\phantom\dagger}_{R{\bf k}\sigma} + \sum_{\sigma} \epsilon_h h^{\dagger}_\sigma h^{\phantom\dagger}_\sigma + \sum_{\sigma} \epsilon_d d^{\dagger}_\sigma d^{\phantom\dagger}_\sigma + U d^{\dagger}_\uparrow
 d^{\dagger}_\downarrow d^{\phantom\dagger}_\downarrow d^{\phantom\dagger}_\uparrow \nonumber\\
  && +\sum_\sigma \left[ t_{Lh} h^{\dagger}_\sigma \psi^{\phantom\dagger}_{L\sigma}({\bf 0})   + t_{Ld} d^{\dagger}_\sigma \psi^{\phantom\dagger}_{L\sigma}({\bf 0}) + t_{Rh}  \psi^{\dagger}_{R\sigma}({\bf 0}) h^{\phantom\dagger}_\sigma 
  + t_{Rd}  \psi^{\dagger}_{R\sigma}({\bf 0}) d^{\phantom\dagger}_\sigma + \mathrm{h.c.} \right]
\label{ModHam}
\end{eqnarray}
Here, electrons of wavevector $\mathbf{k}$ in tip and substrate are annihilated by $\psi^{\phantom\dagger}_{L{\bf k}\sigma}$ and $\psi^{\phantom\dagger}_{R{\bf k}\sigma}$, respectively. The operators $\psi^{\phantom\dagger}_{L\sigma}({\bf 0})$ and $\psi^{\phantom\dagger}_{R\sigma}({\bf 0})$ annihilate electrons in tip and substrate at the location of the tunnel junction to the molecular adsorbate. For the tip, the origin $\mathbf{0}$ is measured relative to a coordinate system attached to the tip. 

The tunneling matrix elements  $t_{Lh}$ and $t_{Ld}$ between tip and molecule depend on the position $\mathbf{R}$ of the tip relative to the molecule, and reflect the wavefunctions of the molecular orbitals. In particular, $t_{Lh}(\mathbf{R})$ changes sign at the nodal planes of the HOMO. There is no evidence in the data for nodal planes in the spin-carrying orbitals, so that we assume that $t_{Ld}(\mathbf{R})$ has a fixed sign. The tunneling amplitudes $t_{Rh}$ and $t_{Rd}$ between molecule and substrate are independent of tip position and fixed by the adsorption geometry of the molecule on the substrate. We finally note that the tunneling amplitudes $t_{Lh}$, $t_{Ld}$, $t_{Rh}$, and $t_{Rd}$ can be chosen as real by time reversal symmetry. 

\subsection{Kondo resonance}

\subsubsection{Fano lineshape of Kondo resonance}

Fano lineshapes emerge from the interference of a resonant channel with a rapid dependence of the scattering amplitude on energy and a nonresonant channel for which the energy dependence can be neglected on the scale of the broadening of the resonance. In the context of adatom experiments, the nonresonant channel is frequently identified with direct tunneling into the substrate which then interferes with a resonant scattering channel via the adatom which results from the Kondo effect \cite{SMadhavan1998,SSchiller2000}. For molecular adsorbates, tunneling between tip and substrate is presumably mediated by molecular orbitals as long as the tip is positioned above the molecule \cite{SAnderson1966}. In keeping with the model Hamiltonian in Eq.\ (\ref{ModHam}), we therefore neglect direct tunneling between tip and substrate and show that the Fano lineshape is naturally explained by interference between tunneling paths associated with two molecular orbitals. A recent study argued that similar processes are also relevant for Kondo resonances induced by magnetic adatoms \cite{SFrank2015}.

Assuming weak tunneling between tip and molecular adsorbate, the differential conductance can be written as 
\begin{equation}
  \frac{\mathrm{d} I}{\mathrm{d}V} = \frac{2e^2}{h} 2\pi\nu_L \rho_{rr^\dagger}(eV),
\label{DiffConSpecDen}
\end{equation}
where $\nu_L$ is the electronic density of states of the tip and $\rho_{rr^\dagger}(\omega)$ denotes the relevant spectral function of the substrate,
\begin{equation}
  \rho_{rr^\dagger}(\omega) = -2\mathrm{Im} G_{rr^\dagger}(\omega+i\eta)
\end{equation}
($\eta$ denotes a positive infinitesimal) with 
\begin{equation}
   r^\dagger  = t_{Lh} h^\dagger + t_{Ld} d^\dagger 
\end{equation}
and $G_{rr^\dagger}(\omega+i\eta)$ denoting the Fourier transform of the retarded Green function
\begin{equation}
  G_{rr^\dagger}(t,t') = -i\theta(t-t') \langle [r(t),r^\dagger(t')]\rangle.
\label{GreenF}
\end{equation}
Since the spectral function is proportional to a unit matrix in spin space, we have dropped spin indices. We can then write
\begin{eqnarray}
   \frac{\mathrm{d} I}{\mathrm{d}V} = \frac{2e^2}{h} 4\pi^2\nu_L \, 
 \frac{-1}{\pi}\mathrm{Im} \left\{ t_{Lh} G_{hh^\dagger}(eV) t_{Lh} + t_{Lh} G_{hd^\dagger}(eV) t_{Ld} + t_{Ld} G_{dh^\dagger}(eV) t_{Lh} + t_{Ld} G_{dd^\dagger}(eV) t_{Ld}
\right\},
\label{DiffConInt}
\end{eqnarray}
where the Green functions are defined by analogy with Eq.\ (\ref{GreenF}) and the positive infinitesimal is left implicit. 

As a consequence of the Kondo effect, the Green function $G_{dd^\dagger}(\omega+i\eta)$ of the spin-carrying orbital exhibits a sharp resonance at the Fermi energy. We therefore relate the remaining Green functions in Eq.\ (\ref{DiffConInt}) to $G_{dd^\dagger}(\omega+i\eta)$. This yields (see Ref.\ \cite{SSchiller2000} for a similar expression in the case that the parallel channel originates from direct tunneling into the substrate)
\begin{eqnarray}
   \frac{\mathrm{d} I}{\mathrm{d}V} &=& \frac{2e^2}{h} 4\pi^2\nu_L \frac{-1}{\pi}\mathrm{Im} \left\{ t_{Lh} \tilde g_{hh^\dagger}(eV) t_{Lh} \right.\nonumber\\
  &+& \left. \left[ t_{Ld} + t_{Lh} \tilde g_{hh^\dagger}(eV) t_{Rh} g_{R R^\dagger}(eV) t_{Rd}] G_{dd^\dagger}(eV) [ t_{Ld} + t_{Rd} g_{R 
   R^\dagger}(eV) t_{Rh} \tilde g_{hh^\dagger}(eV) t_{Lh} \right] \right\}.\,\,\,\,\,\,\,\,\,\,\,
\label{DiffConGreen}
\end{eqnarray}
Here, a capital $G$ denotes a fully dressed Green function, a lower-case $g$ denotes a bare Green function ($g$) or a Green function including a subset of self-energy contributions ($\tilde g$). We sketch the derivation of Eq.\ (\ref{DiffConGreen}) in the subsequent section. Since it can also be readily understood physically, we proceed first with a qualitative justification and focus on how it predicts a Fano lineshape for the Kondo resonance. The first term within the curly brackets describes tunneling from tip to substrate via the HOMO in the absence of the $d$ orbital. Correspondingly, this term involves the Green function
\begin{equation}
  \tilde g_{hh^\dagger} (\omega) = \frac{1}{\omega - \epsilon_h - t_{Rh} g_{R R^\dagger}(\omega) t_{Rh} } = \frac{1}{\omega - \epsilon_h +  i\pi (t_{Rh})^2 \nu_R}
\end{equation}
of the HOMO including the self energy correction due to tunneling into the bare substrate. The latter enters through the Green function $g_{R R^\dagger}(\omega)$ of the bare substrate in the absence of the molecular adsorbate. Neglecting the spatial extent of the molecule for simplicity, this Green function enters only with equal spatial indices of its two electron operators, and we have
\begin{equation}
    g_{R R^\dagger}(\omega) = -i\pi \nu_R
\end{equation}
in terms of the electronic density of states $\nu_R$ of the bare substrate. The second term in the curly bracket collects all contributions which proceed via the $d$ orbital and are thus sensitive to the Kondo resonance. The contribution involving $t_{Ld}^2$ describes direct tunneling from the tip into the $d$ orbital, dressed by its coupling to the remainder of the system. Similarly, the term proportional to $t_{Lh}^2$ describes a process in which an electron tunnels from the tip into the dressed $d$ orbital via a multistep process involving the HOMO and the substrate. Finally, the terms involving $t_{Ld}t_{Lh}$ describe contributions which arise from interference of the previous two tunneling paths.  

To bring out the essential physics of the Fano lineshape, we note that the two terms in square brackets in Eq.\ (\ref{DiffConGreen}) are identical and complex. Defining the polar representation 
\begin{equation}
  \sqrt{\mathcal{A}}e^{i\phi/2} = t_d + t_{Lh} \tilde g_{hh^\dagger}(eV) t_{Rh} g_{R R^\dagger}(eV) t_{Rd},
\label{pol}
\end{equation}
Eq.\ (\ref{DiffConGreen}) becomes
\begin{eqnarray}
   \frac{\mathrm{d} I}{\mathrm{d}V} = \frac{2e^2}{h} 4\pi^2\nu_L \,\frac{-1}{\pi}\mathrm{Im} \left\{ t_{Lh} \tilde g_{hh^\dagger}(eV) t_{Lh} 
  + \mathcal{A}e^{i\phi} G_{dd^\dagger}(eV)  \right\}.
\label{DiffConA}
\end{eqnarray}
We evaluate Eqs.\ (\ref{pol}) and (\ref{DiffConA}) by keeping only the leading nonvanishing contributions of  $\tilde g_{hh^\dagger} (\omega)$ in the limit in which the broadening of the HOMO is small compared to its energy $\epsilon_h$ measured from the Fermi energy. (It should be evident that it is straight-forward to go beyond this approximation, but this is unnecessary for our purposes.)  With this approximation, $\tilde g_{hh^\dagger} (eV)\simeq -i\pi (t_{Rh})^2\nu_R/\epsilon_h^2$ in the first term in curly bracket in Eq.\ (\ref{DiffConA}) and $\tilde g_{hh^\dagger} (eV)\simeq -1/\epsilon_h$ in Eq.\ (\ref{pol}). This yields
\begin{equation}
   \sqrt{\mathcal{A}}e^{i\phi/2} =   t_d + i\pi \frac{t_{Lh}  t_{Rh}}{\epsilon_h} \nu_R t_{Rd}.
\end{equation}
Since the imaginary part of the right-hand side is proportional to the tunneling amplitude $t_{Lh}$ into the HOMO of the molecule, the phase $\phi$ passes through zero and changes sign at the nodal plane.  

Within a simple, but useful approximation, we can express the $d$-orbital Green function in the vicinity of the Fermi energy as \cite{SUjsaghy2000}
\begin{equation}
   G_{dd^\dagger}(eV) = \frac{Z_K}{eV - \epsilon_K +i T_K}, 
\end{equation}
where $Z_K$ denotes the intrinsic strength of the Kondo resonance, $\epsilon_K\simeq 0$ its energy, and $T_K$ the Kondo temperature. The shape of the Kondo resonance is frequently fitted more accurately by the Frota function \cite{SFrota1992}
\begin{equation}
   G_{dd^\dagger}(eV) = \frac{-i Z_K}{T_K} \sqrt{\frac{iT_K}{eV - \epsilon_K +i T_K}}.
\end{equation}
Inserting this expression into Eq.\ (\ref{DiffConA}), we obtain 
\begin{eqnarray}
  \left. \frac{\mathrm{d} I}{\mathrm{d}V}\right|_\mathrm{res} = \frac{2e^2}{h} 4\pi^2\nu_L \mathrm{Re} \left\{\mathcal{A}e^{i\phi} \frac{ Z_K}{\pi T_K} \sqrt{\frac{iT_K}{eV - \epsilon_K +i T_K}}  \right\}.
\label{DiffConRes}
\end{eqnarray}
This expression has the same form as Eq.\ (1) of the main text. It also shows that the sign change of the HOMO wavefunction changes the sense of asymmetry of the Kondo resonance, as we observe experimentally. 

\subsubsection{Green-function expression (\ref{DiffConGreen})}

In this section, we sketch the derivation of Eq.\ (\ref{DiffConGreen}) from Eq.\ (\ref{DiffConInt}). This requires one to relate the various Green functions entering Eq.\ (\ref{DiffConInt}) to $G_{dd^\dagger}$. Since interactions enter only on the $d$ orbital, this can be done with the help of the equations of motions of the Green functions. One readily finds the Dyson equations
\begin{eqnarray}
   G_{hh^\dagger} &=& g_{hh^\dagger} + g_{hh^\dagger} t_{Rh} G_{Rh^\dagger} \label{1} \\
   G_{hR^\dagger} &=&  g_{hh^\dagger} t_{Rh} G_{RR^\dagger}    \label{2}\\
   G_{hd^\dagger} &=&  g_{hh^\dagger} t_{Rh} G_{Rd^\dagger}    \label{3}\\
   G_{Rh^\dagger} &=&  g_{RR^\dagger} t_{Rh} G_{hh^\dagger} + g_{RR^\dagger} t_{Rd} G_{dh^\dagger} \label{4}   \\
   G_{RR^\dagger} &=&  g_{RR^\dagger} + g_{RR^\dagger} t_{Rh} G_{hR^\dagger} + g_{RR^\dagger} t_{Rd} G_{dR^\dagger} \label{5}   \\
   G_{Rd^\dagger} &=&  g_{RR^\dagger} t_{Rh} G_{hd^\dagger} + g_{RR^\dagger} t_{Rd} G_{dd^\dagger}  \label{6}  \\
   G_{Rh^\dagger} &=&  G_{RR^\dagger} t_{Rh} g_{hh^\dagger}   \label{7} \\
   G_{dh^\dagger} &=&  G_{dR^\dagger} t_{Rh} g_{hh^\dagger}   \label{8} \\
   G_{dR^\dagger} &=&  G_{dh^\dagger} t_{Rh} g_{RR^\dagger} + G_{dd^\dagger} t_{Rd} g_{RR^\dagger}. \label{9}   
\end{eqnarray}
Here, $g_{hh^\dagger}(\omega)= 1/(\omega - \epsilon_h + i\eta)$ denotes the Green function of the uncoupled HOMO. The first six expressions result from equations of motions involving time derivatives with respects to the first time argument of the Green function. The last three identities follow from equations of motions with respect to the second time argument. 

We first insert Eq.\ (\ref{9}) into Eq.\ (\ref{8}) and solve for $G_{dh^\dagger}$. This yields
\begin{equation}
   G_{dh^\dagger} = G_{dd^\dagger} t_{Rd} g_{RR^\dagger} t_{Rh} g_{hh^\dagger} \frac{1}{1- t_{Rh} g_{RR^\dagger} t_{Rh} g_{hh^\dagger}}
\end{equation}
and thus
\begin{equation}
   G_{dh^\dagger} = G_{dd^\dagger} t_{Rd} g_{RR^\dagger} t_{Rh} \tilde g_{hh^\dagger}. 
\label{dsdagger}
\end{equation}
Similarly, inserting Eq.\ (\ref{6}) into Eq.\ (\ref{3}) and solving for $G_{hd^\dagger}$, we find  
\begin{equation}
   G_{hd^\dagger} =  \frac{1}{1-g_{hh^\dagger} t_{Rh} g_{RR^\dagger} t_{Rh} } g_{hh^\dagger} t_{Rh} g_{RR^\dagger} t_{Rd} G_{dd^\dagger}
\end{equation}
and thus
\begin{equation}
   G_{hd^\dagger} = \tilde g_{hh^\dagger} t_{Rh} g_{RR^\dagger} t_{Rd} G_{dd^\dagger}. 
\label{sddagger}
\end{equation}
Finally, we insert Eq.\ (\ref{4}) into Eq.\ (\ref{1}) and solve for $G_{hh^\dagger}$. This yields
\begin{equation}
   G_{hh^\dagger} = \frac{1}{1- g_{hh^\dagger} t_{Rh} g_{RR^\dagger} t_{Rh} } g_{hh^\dagger} (1 + t_{Rh}  g_{RR^\dagger}  t_{Rd} G_{dh^\dagger})
\end{equation}
and thus
\begin{equation}
   G_{hh^\dagger} =  \tilde g_{hh^\dagger}  + \tilde g_{hh^\dagger} t_{Rh}  g_{RR^\dagger}  t_{Rd}  G_{dh^\dagger}.
\end{equation}
Inserting Eq.\ (\ref{dsdagger}), we obtain
\begin{equation}
   G_{hh^\dagger} =  \tilde g_{hh^\dagger}  + \tilde g_{hh^\dagger} t_{Rh}  g_{RR^\dagger}  t_{Rd}  G_{dd^\dagger} t_{Rd} g_{RR^\dagger} 
    t_{Rh} \tilde g_{hh^\dagger}.
\label{ssdagger}
\end{equation}
We can now insert Eqs.\ (\ref{dsdagger}), (\ref{sddagger}), and (\ref{ssdagger}) into Eq.\ (\ref{DiffConInt}) and obtain Eq.\ (\ref{DiffConGreen}).

\subsection{YSR resonances}

\subsubsection{Schrieffer-Wolff transformation}

To understand the asymmetry of the YSR resonances at positive and negative bias voltages, it is convenient to eliminate the spin-carrying $d$ orbital as well as the HOMO by a Schrieffer-Wolff transformation and to obtain an effective low-energy Hamiltonian. 

The molecular $d$ orbital leads to exchange and potential scattering terms, 
\begin{equation}
          H_d = \sum_{\alpha\alpha'} J^{(d)}_{\alpha\alpha'}(\mathbf{R}) \mathbf{S}\cdot \psi_\alpha^\dagger(\mathbf{0})\boldsymbol {\sigma} \psi^{\phantom\dagger}_{\alpha'}(\mathbf{0}) + \sum_{\alpha\alpha'}  V^{(d)}_{\alpha\alpha'}(\mathbf{R})  \psi_\alpha^\dagger(\mathbf{0}) \psi^{\phantom\dagger}_{\alpha'}(\mathbf{0})
\end{equation}
with the impurity-spin operator ${\bf S}$, the vector $\boldsymbol\sigma$ of Pauli matrices, and the spinor of electron field operators $\psi_{\alpha}(\mathbf{0})= [\psi_{\alpha,\uparrow}(\mathbf{0}),\psi_{\alpha,\downarrow}(\mathbf{0})]^T$ in the tip ($\alpha=L$) and the substrate ($\alpha=R$; with $\mathbf{0}$ denoting the adsorption position of the molecule). In contrast, the nonmagnetic HOMO gives rise to a potential scattering term only,  
\begin{equation}
          H_{h} = \sum_{\alpha\alpha'}   V^{(h)}_{\alpha\alpha'}(\mathbf{R})  \psi_\alpha^\dagger(\mathbf{0}) \psi^{\phantom\dagger}_{\alpha'}(\mathbf{0}).
\end{equation}
The Schrieffer-Wolff transformation yields 
\begin{equation}
  J^{(d)}_{\alpha\alpha'}(\mathbf{R}) = t_{\alpha d}^{\phantom *} t_{\alpha^\prime d}^* \left[ \frac{1}{\epsilon_d+U} - \frac{1}{\epsilon_d}\right] 
\end{equation}
for the exchange coupling and 
\begin{equation}
  V^{(d)}_{\alpha\alpha'}(\mathbf{R}) = t_{\alpha d}^{\phantom *} t_{\alpha^\prime d}^* \left[ \frac{1}{\epsilon_d+U} + \frac{1}{\epsilon_d}\right] 
\end{equation}
for the amplitude of potential scattering from the molecular $d$ orbital. Similarly, the HOMO results in the amplitude
\begin{equation}
  V^{(h)}_{\alpha\alpha'}(\mathbf{R}) = t_{\alpha h}^{\phantom *} t_{\alpha^\prime h}^*  \frac{1}{\epsilon_h}
\label{Vh}
\end{equation}
of potential scattering. 

In experiment, the tip-molecule coupling is much weaker than the molecule-substrate coupling. We thus neglect the exchange and potential-scattering processes which only involve the tip (i.e., we set $J^{(d)}_{LL} = V^{(d)}_{LL}=V^{(h)}_{LL}=0$) and treat terms which scatter electrons between tip and substrate perturbatively. In leading order, the molecule is then coupled to the substrate through $J^{(d)}_{RR}$ and $V^{(d)}_{RR}$ as well as $V^{(h)}_{RR}$. The strong Kondo correlations arise as the exchange coupling flows to strong coupling at low energies, and we can thus neglect $V^{(d/h)}_{RR}$ relative to $J^{(d)}_{RR}$ in discussing low-bias tunneling. For the same reason, scattering between tip and substrate via the spin-carrying orbital is dominated by the exchange term, and we neglect $V^{(d)}_{RL}$ and $V^{(d)}_{LR}$.

As a result of these considerations, we obtain the effective Hamiltonian
\begin{equation}
  H = H_0 + H_1
\end{equation}
with unperturbed Hamiltonian 
\begin{equation}
   H_0 =  \sum_\alpha H_\alpha + J^{(\rm d)}_{RR}(\mathbf{R}) \mathbf{S}\cdot \psi_R^\dagger(\mathbf{0})\boldsymbol{\sigma} \psi_{R}(\mathbf{0}) .
\end{equation}
and tunneling perturbation 
\begin{equation}
   H_1 =  \psi_R^\dagger(\mathbf{0})\left[ J^{(\rm d)}_{RL}(\mathbf{R}) \mathbf{S}\cdot\boldsymbol{\sigma} +  V^{(\rm h)}_{RL}(\mathbf{R}) \right] \psi_{L}(\mathbf{0}) + \mathrm{h.c.} .
\end{equation}
As time-reversal symmetry implies that all tunneling amplitudes can be chosen real, this also holds true for $J^{(\rm d)}_{\alpha\alpha'}(\mathbf{R})$ and $V^{(\rm h)}_{\alpha\alpha'}(\mathbf{R})$. The Kondo correlations as well as the YSR states are induced by the exchange coupling between molecular spin $\mathbf{S}$ and substrate included in $H_0$. Tunneling between tip and substrate proceeds via two separate channels involving the molecular $d$ orbital and the HOMO, respectively. While the second tunneling channel is independent of electron spin, the first is spin dependent. According to Eq.\ (\ref{Vh}), the sign change of the HOMO wavefunction at the nodal plane introduces a sign change of the potential scattering amplitude $V^{(h)}(\mathbf{R})$. 

 \subsubsection{Asymmetry of YSR resonances}

We now derive the tunneling current into the YSR state by Fermi's golden rule. At large tip distances, tunneling between tip and substrate is rate limited by the tunneling between tip and YSR state \cite{SRuby2015}. The resulting subgap excitation (i.e., a quasiparticle occupying the YSR state) relaxes into the quasiparticle continuum by inelastic processes which are more frequent than tunneling. As a consequence of the strong Kondo correlations, the YSR state originates from strong exchange coupling $J_{RR}$ between molecular orbital and substrate, with potential scattering $V_{RR}$ being a small perturbation. In this case, one expects that the electron and hole wavefunctions of the YSR states are approximately equal. For instance, it is well known that the asymmetry between electron and hole wavefunctions is due to nonzero potential scattering when modeling the impurity as a classical spin. 

Motivated by the experiment, we assume that the YSR state is on the weak-coupling (unscreened) side of the quantum phase transition between doublet (unscreened) and singlet (screened) ground states. (As mentioned in the main text, this assumption is made for definiteness. Our conclusions would not change for a YSR state in the strong-coupling state and the following arguments can be readily adapted to this case.) In this case, the substrate superconductor is in a fully paired ground state $|\mathrm{even}\rangle$ with even fermion parity, and tunneling is into the excited odd-fermion parity state $|\mathrm{odd}\rangle = \gamma^\dagger_\epsilon |\mathrm{even}\rangle$ with a quasiparticle occupying the YSR state (with Bogoliubov operator $\gamma_\epsilon$). For simplicity, we assume a classical-spin model for the impurity in the following, i.e., we effectively keep only the exchange coupling to $S_z$. 
This suffices to elucidate the emergence of the asymmetry between the bias directions due to the presence of two parallel tunneling channels. As this asymmetry is clearly associated with the different behaviors of exchange and potential scattering under time reversal of the substrate Hamiltonian (for a given impurity spin), the underlying mechanism is expected to be robust when extending the theory to include the full quantum nature of the impurity spin. Developing such a theory would be an interesting problem for future research, but is beyond the scope of the present paper.

To compute the tunneling rate between superconducting tip and substrate, we expand the electron operator of the substrate into Bogoliubov operators. At subgap bias voltages, we only need to retain the contribution of the YSR state. We take the impurity spin as polarized in the positive $z$-direction, so that the YSR bound state is polarized along the negative $z$-direction. We can then write the substrate electron operators as 
\begin{equation}
  \psi_{R,\downarrow}({\bf r}) = u_\epsilon({\bf r}) \gamma_\epsilon+\ldots
\label{down}
\end{equation} 
and 
\begin{equation}
  \psi_{R,\uparrow}({\bf r}) = -v^*_\epsilon({\bf r})\gamma_\epsilon^\dagger + \ldots,
\label{up}
\end{equation} 
with the ellipses referring to the contributions of above-gap quasiparticles. Here, we have introduced the electron and hole wavefunctions $u_\epsilon({\bf r})$ and $v_\epsilon({\bf r})$ of the YSR state. These equations reflect that due to the spin polarization of the YSR state, removing a quasiparticle is associated with annihilating a down-spin electron. Conversely, removing an up-spin electron from the ground state necessarily breaks a Cooper pair and its spin-down electron can subsequently occupy the YSR state. 

First consider tunneling from the tip into the YSR state of energy $\epsilon_s$ at positive bias voltages. Fermi's golden rule gives the tunnel current
\begin{equation}
   I|_+ =  \frac{2\pi e}{\hbar} \sum_{{\bf k},\sigma} |\langle \mathrm{odd};E_{Lk}|H_1|\mathrm{even};\mathrm{BCS}_L\rangle|^2
      \delta(eV - E_{Lk}-\epsilon_s),
\end{equation}
where $|\mathrm{BCS}_L\rangle$ and $|E_{Lk}\rangle$ denote the BCS ground state and a state with a single Bogoliubov quasiparticle of energy $E_{L{\bf k}}$ in the tip. According to Eqs.\ (\ref{down}) and (\ref{up}) and in accordance with the spin polarization of the YSR state, tunneling into the YSR state is due to spin-down electrons,
\begin{eqnarray}
   I|_+ &=&  \frac{2\pi e}{\hbar} \sum_{{\bf k}} |\langle \mathrm{odd};E_{Lk}|\psi_{R\downarrow}^\dagger(\mathbf{0})\left[ -J^{(\rm d)}_{RL}(\mathbf{R}) {S} +  V^{(\rm h)}_{RL}(\mathbf{R}) \right] \psi_{L\downarrow}(\mathbf{0})|\mathrm{even};\mathrm{BCS}_L\rangle|^2
      \delta(eV - E_{Lk}-\epsilon_s) \nonumber\\
 &=&  \frac{2\pi e}{\hbar} |u_\epsilon({\bf 0})|^2 \left[ -J^{(\rm d)}_{RL}(\mathbf{R}) {S} +  V^{(\rm h)}_{RL}(\mathbf{R}) \right]^2\sum_{{\bf k}} |\langle E_{Lk}| \psi_{L\downarrow}(\mathbf{0})|\mathrm{BCS}_L\rangle|^2
      \delta(eV - E_{Lk}-\epsilon_s).
\end{eqnarray}
The sum over ${\bf k}$ can be identified with the superconducting density of states of the tip, 
\begin{equation}
    \nu_{L,\mathrm{SC}}(E) =  \nu_L \frac{|E|}{\sqrt{E^2 - \Delta^2 }}\theta(\Delta - |E|),
\end{equation}
at energy $E = eV-\epsilon_s$, so that we obtain
\begin{equation}
   I|_+ = \frac{2\pi e}{\hbar} |u_\epsilon({\bf 0})|^2 \left[ -J^{(\rm d)}_{RL}(\mathbf{R}) {S} +  V^{(\rm h)}_{RL}(\mathbf{R}) \right]^2
 \nu_{L,\mathrm{SC}}(eV-\epsilon_s).
\label{+}
\end{equation}
Importantly, the current involves interference between an exchange contribution associated with the molecular $d$ orbital and a potential scattering contribution associated with the HOMO. The relative sign with which these contributions contribute directly reflects the fact that the tunneling current from tip to substrate is carried by spin-down electrons.  

We will now see that the relative sign between these contributions is reversed at negative bias voltages, as in this case, current is carried by spin-up electrons. This can be seen from Eqs.\ (\ref{down}) and (\ref{up}). Physically, tunneling of an electron from the substrate into the tip breaks a Cooper pair in the even-parity ground state of the substrate. The spin-down electron of the Cooper pair occupies the YSR state, while the spin-up electron tunnels into the tip. This leads to
\begin{eqnarray}
   I|_- &=&  \frac{2\pi e}{\hbar} \sum_{{\bf k}} |\langle \mathrm{odd};E_{Lk}|\psi_{L\uparrow}^\dagger(\mathbf{0})\left[ J^{(\rm d)}_{RL}(\mathbf{R}) {S} +  V^{(\rm h)}_{RL}(\mathbf{R}) \right] \psi_{R\uparrow}(\mathbf{0})|\mathrm{even};\mathrm{BCS}_L\rangle|^2
      \delta(-eV - E_{Lk}-\epsilon_s) \nonumber\\
 &=&  \frac{2\pi e}{\hbar} |v_\epsilon({\bf 0})|^2 \left[ J^{(\rm d)}_{RL}(\mathbf{R}) {S} +  V^{(\rm h)}_{RL}(\mathbf{R}) \right]^2 \nu_{L,\mathrm{SC}}(e|V|-\epsilon_s)
\label{-}
\end{eqnarray}
for the YSR resonance negative bias. 

Comparing Eqs.\ (\ref{+}) and (\ref{-}), we see  that the relative signs of the exchange and potential-scattering contributions to the tunneling current differ between positive and negative bias voltages,
\begin{equation}
   \left. \frac{\mathrm{d} I}{\mathrm{d}V}\right|_\pm \propto \left[ \mp J^{(\rm d)}_{RL}(\mathbf{R}) {S} +  V^{(\rm h)}_{RL}(\mathbf{R}) \right]^2,
\end{equation}
with the other contributions being essentially independent of the bias direction. Indeed, the electron and hole wavefunctions of the YSR state should be approximately equal, $|u_\epsilon({\bf 0})|\simeq|v_\epsilon({\bf 0})|$, due to the dominance of exchange coupling to the substrate implied by the well-developed Kondo correlations observed in experiment. We thus conclude that the YSR resonances should be symmetric between the bias directions when tunneling into the nodal plane of the HOMO, and have opposite asymmetries on the two sides of the nodal plane as a result of the sign change of  $V^{(\rm h)}_{RL}(\mathbf{R})$.

\end{document}